# $Zn/Cu_xTi_y$体系固态反应周期层片型结构

巩宇 [1,2,3], 陈永翀[1,*], 张艳萍[1,3], CSERHÁTI Csaba[4,*], CSIK Attila[5]

（[1]Energy Storage Technology Research Group, Institute of Electrical Engineering, Chinese Academy of Science, Beijing 100190, China; [2]University of Chinese Academy of Sciences, Beijing 100049, China; [3]Beijing HAWAGA Power Storage Technology Company Ltd., Beijing 100085，China; [4]Department of Solid State Physics, University of Debrecen, H-4010 Debrecen, Hungary; [5]Institute for Nuclear Research, Hungarian Academy of Sciences, H-4001 Debrecen, Hungary）

单击输入作者名,两名字间用空格间隔，不同单位用上标 1)，2）等区分

1）中国科学院电工研究所, 北京 100190

2）北京好风光储能技术有限公司，北京，100085

**摘 要** 采用液接法制备 $Zn/Cu_xTi_y$ 扩散偶, 利用电子显微镜和能谱分析仪等手段对扩散偶在 390℃保温不同时间后的反应区进行分析。除了之前文献报道过的 $Zn/CuTi_2$ 和 Zn/CuTi 体系以外，我们新发现了其它 7 个扩散反应体系能够形成周期层片型结构，分别是 $Zn/Cu_9Ti$，$Zn/Cu_4Ti$，$Zn/Cu_2Ti$，$Zn/Cu_7Ti_3$，$Zn/Cu_3Ti_2$，$Zn/Cu_{11}Ti_9$，$Zn/Cu_9Ti_{11}$ 。对扩散偶反应区域分别进行金相抛光及原位断面的高倍扫描观测，发现 $Zn/Cu_xTi_y$ 体系靠近反应区前沿的周期层片结构由单相（$CuZn_2$）和双相（$CuZn_2$+$TiZn_3$）交替构成，而且周期层片结构中片层厚度与 Cu-Ti 合金成分有关：Cu-Ti 合金中 Cu 原子含量越高，对应反应区的周期层片厚度越小，上述实验结果符合扩散应力模型的预测。

# INVESTIGATION OF PERIODIC-LAYERED STRUCTURE IN $Zn/Cu_xTi_y$ SYSTEMS

*GONG Yu* [1,2,3], *CHEN Yongchong*[1,*], ZHANG Yanping[1,3], *CSERHÁTI Csaba*[4,*], *CSIK Attila*[5]

（[1]Energy Storage Technology Research Group, Institute of Electrical Engineering, Chinese Academy of Science, Beijing 100190, China; [2]University of Chinese Academy of Sciences, Beijing 100049, China; [3]Beijing HAWAGA Power Storage Technology Company Ltd., Beijing 100085，China; [4]Department of Solid State Physics, University of Debrecen, H-4010 Debrecen, Hungary; [5]Institute for Nuclear Research, Hungarian Academy of Sciences, H-4001 Debrecen, Hungary）

**ABSTRACT**

Diffusion couples $Zn/Cu_xTi_y$ were prepared by the melting contact method and then annealed at 663K for various times. Using scanning electron microscopy (SEM) with energy dispersive X-ray spectroscopy (EDS), we discovered 7 new systems, i.e. $Zn/Cu_9Ti$, $Zn/Cu_4Ti$, $Zn/Cu_2Ti$, $Zn/Cu_7Ti_3$, $Zn/Cu_3Ti_2$, $Zn/Cu_{11}Ti_9$ and $Zn/Cu_9Ti_{11}$, which can form the interesting periodic-layered structure within the reaction zones. By the traditional metallographic treatment and the in-situ section observation method respectively, it was confirmed that the periodic-layered structure is really composed of the $CuZn_2$ singer-phase and the ($CuZn_2$+$TiZn_3$) two-phase layers distributing alternately within the reaction front area. Furthermore, the thickness of the periodic layers relates to the composition of Cu-Ti substrates: the higher content of Cu atoms in Cu-Ti alloy substrates, the thinner the layers will be. The experimental results are in accordance to the prediction of the diffusion-induced stresses model.

*输入基金中文名称及编号
收到初稿日期：200*—**—**，收到修改稿日期：200*—**—**
作者简介：巩宇，男，汉族，1989 年生，研究生

固态反应周期层片型结构是一类高度规则的微纳米级自生成复合多层膜结构，膜层界面结合良好，是未来功能薄膜材料制备技术的发展方向之一。迄今发现能够形成周期层片结构的固态反应体系有：Zn/$Fe_3Si$[1]，Zn/$Co_2Si$[2,3]，Zn/$Ni_3Si$,Ni/SiC, Mg/$Ni_{50}Co_{20}Fe_{30}$[4]，Pt/SiC，Co/SiC[5,6]，Zn/$Ni_3Si_2$[7], Mg/$SiO_2$[8],Zn/$Ni_3Si$[9],Al/$U_{10}$Mo[10],Al/(Ni,W)[11],Ni/A-$u_{12}$Ge[12]，以及最近发现的 Zn/$CuTi_2$[13]和 Zn/CuTi[14]。

历史上提出过两类不同的理论模型用于解释固态反应周期层片型结构的形成原因：一类是热力学失稳机制，另一类是动力学失稳机制[15,16]。

热力学失稳机理由 Dunaev 和 Zver'kov[4]首次提出，Kao 和 Chang[17]建立了定量模型。Gutman I, 和 Gotman I 利用此机理尝试解释 Mg/$SiO_2$[18-20]体系周期层片型结构的形成。热力学失稳机理认为，基体 A 和基体 B 的反应界面首先形成基体 A| α | β |基体 B 的简单层片结构，随着反应扩散的进行，当形成 α 相的元素浓度在 β 相与基体 B 的接触界面达到过饱和时，α 相将形核析出，并与紧邻基体 B 的 β 相同时长大；当反应进行到一定程度时，新的 α 相又在 β 相与基体 B 的接触界面附近形核析出。如此反复，三元扩散偶的反应扩散区就形成了单相 α 与单相 β 交替构成的周期层片结构（基体 A | α | β | α |···| α | β | 基体 B）。

然而，实验证明，目前发现的所有固态反应周期层片型结构都不是由单相 α 与单相 β 交替构成的，而是由单相 α 与双相（α+β）交替构成[13,14]。这意味着在反应扩散的过程中，α 相和 β 相其实一直以束集交联结构（aggregate-interwoven structure）的形式同时存在于反应前沿的区域，α 相并不需要在紧邻基体 B 的 β 相内过饱和析出。因此，基于过饱和析出理论的热力学失稳模型不能够解释固态反应周期层片型结构的形成原因。

由于相邻层片间存在相互吻合的形貌特征，Osinski 等[1]在 1982 年最初发现 Zn/$Fe_3Si$ 体系时，曾猜测层片结构的形成可能与反应过程中积累的应力有关，但是不太清楚应力是如何积累的，又是如何促进层片的周期形成。2003 年陈永翀等[16]在固体互扩散生长理论的基础上[21]，建立了一个扩散应力模型（diffusion-induced stresses model），提出形成周期层片结构的动力学失稳机制。在三元扩散偶的反应界面最初生成的是单相 α 和双相（α+β）的层片结构（基体 A | α | α+β |基体 B），在固态反应过程中，如果束集结构中两相的生长速率不同（α 相生长速率快，β 相慢），生长速率慢的 β 相会受到来自快速生长的 α 相的膨胀拉应力，当应力积累到一定程度时 β 相在反应前沿处被撕裂，断裂的缝隙被 α 相生长填充，并逐渐形成一个单相 α 的层片，与与紧邻基体 B 的(α+β)双相束集结构同时长大，厚度增加。如此反复，就形成了基体 A | α | α+β | α |···| α+β | α | α+β |基体 B 的周期层片结构。利用此模型对 Mg/$SiO_2$ 体系的计算模拟结果也能够与实验数据很好的吻合[22]。到目前为止，还没有发现与扩散应力模型相违背的固态反应周期层片型结构体系。

探索新的周期层片结构体系是非常重要的研究工作，这将为以后的复合多层膜结构设计提供更丰富的材料体系选择。最初我们在 Zn/$CuTi_2$ 和 Zn/CuTi 扩散偶中发现了周期层片结构[13,14]，当我们把调查扩展至其它成分的铜钛合金时，又发现了 7 个新的固态反应体系能够形成周期层片机构，在此进行详细报道。

## 1 实验材料及方法

本实验所用 Cu-Ti 合金采用纯度≥99.99wt.%的金属钛片和铜丝按一定的质量比真空熔炼而成，将凝固后未经过均质化处理的合金线切割成 5mm×5mm×3mm 的小块，同时将直径为 8mm，纯度 99.999wt.%的 Zn 棒切割成 1cm 左右，经过超声处理和砂纸打磨将 Cu-Ti 小块和 Zn 棒清洗干净，表面打磨平整。

为保证反应界面结合良好，采用瞬间液接法（melting contact method）制备 Zn/$Cu_xTi_y$ 扩散偶。将制备好的铜钛合金小块与 Zn 块真空密封于石英管中，加热石英管使锌块融化，晃动石英管确保合金被液态锌包裹，静置待 Zn 凝固后迅速把石英管放入 663K 的保温炉中，保温不同时间段后取出空冷。

根据需要采用两种方法制备金相样品，一种方法是将空冷后的扩散偶经过取样、磨制、抛光三步制备金相试样；另一种方法是先将扩散偶预处理，沿一个面小心打磨（如图 1（b）)，磨去外层包裹的

锌层，找到 Cu-Ti 合金小块，然后沿穿过 Cu-Ti 合金的任意一条直线（如图 1(a) 中 AB 方向），从直线两端向合金方向线切割，确保 Cu-Ti 合金保留 2-3mm 连接，然后用机械力将试样掰开（如图 1(b)），对两种金属的反应界面进行原位断面观测。采用扫描电镜（SEM, Zeiss SIGMA）观察扩散偶的反应区形貌，并利用能谱分析仪(EDS, TEAM EDS)分析相区成份。

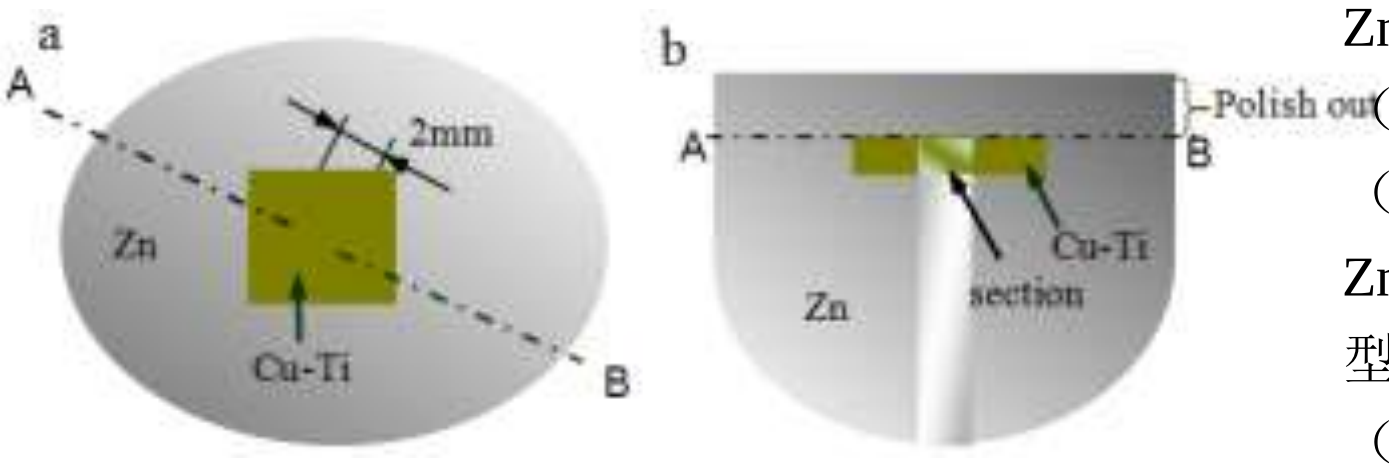

**图 1** 断面观测样品制备示意图

**Fig.1**Diagram of sample making to observe it’s ection.

## 2 实验结果与分析

### 2.1 发现新体系

Cu-Ti 二元相图[23]显示能够稳定存在合金有 CuTi，$Cu_4Ti$，$Cu_2Ti$，$Cu_3Ti_2$，$Cu_4Ti_3$，$CuTi_2$，在合金熔炼时两种金属溶液混合不均匀且后续没有进行足够长时间的退火处理则合金体内会出现成分不均匀。本实验采用的 Cu-Ti 合金未经过均质化处理，采用扫描电镜和能谱分析对扩散偶反应区观测发现：如图 3(a)、(b)所示，Zn-$Cu_3Ti_2$ 扩散偶反应区中能形成周期层片型结构的反应体系除了 Zn/$Cu_3Ti_2$（$Cu_{59.31}$-$Ti_{40.69}$）外，还出现了 Zn/$Cu_7Ti_3$（$Cu_{69.35}Ti_{30.65}$）、Zn/$Cu_{11}Ti_9$（$Cu_{57.18}Ti_{42.82}$）、Zn/CuTi（$Cu_{51.82}Ti_{48.18}$）、Zn/$Cu_2Ti$（$Cu_{68.15}$-$Ti_{31.85}$）；图 4 为 Zn-$Cu_4Ti$ 扩散偶反应区，图中显示能生成周期层片型结构的体系有 Zn/$Cu_4Ti$（$Cu_{79.62}Ti_{20.38}$），Zn/$Cu_9Ti$（$Cu_{89.88}Ti_{10.12}$）；图 5 为 Zn-$Cu_4Ti_3$ 扩散偶反应区，图中显示能生成周期层片型结构的体系有 Zn/$Cu_9Ti_{11}$($Cu_{45.47}Ti_{54.53}$）,Zn/$Cu_{11}Ti_9$（$Cu_{54.66}Ti_{45.35}$)；加上文献[13,14]中报道的 Zn/CuTi 和 Zn/$CuTi_2$，则能生成周期性层片型结构的 Zn/$Cu_xTi_y$ 扩散反应体系数增加到 9 个。

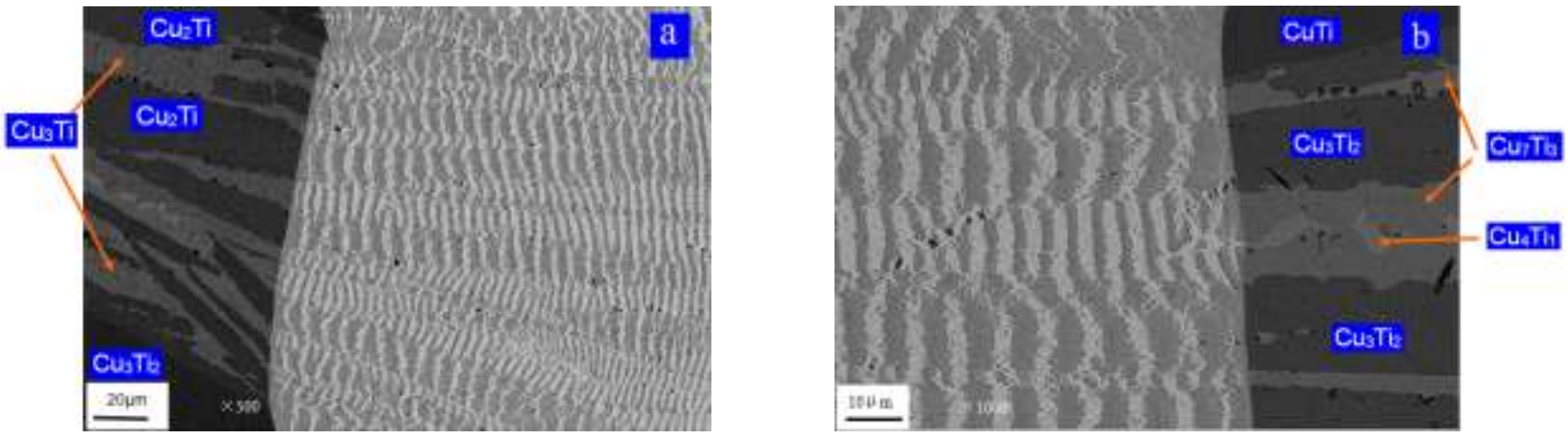

**图 3** Zn/$Cu_3Ti_2$ 体系 663K 真空退火 24h 固态反应周期层片型结构

**Fig.3** SEM images of periodic-layered structure in Zn/$Cu_3Ti_2$ diffusion couple after annealing at 663k for 24 h.

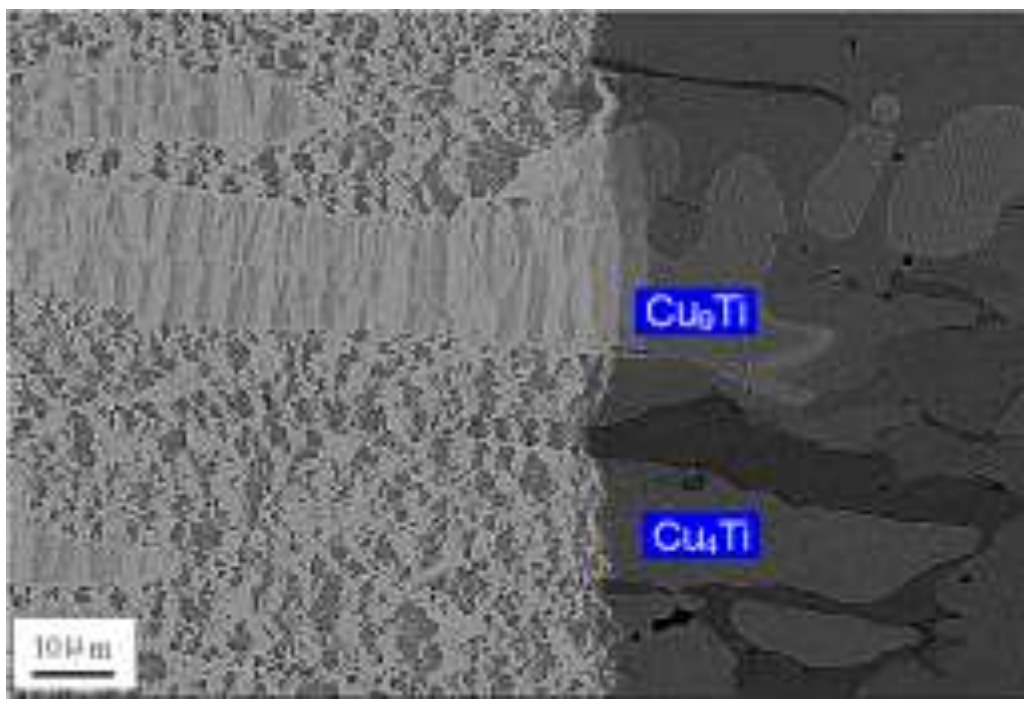

**图 4** Zn/$Cu_4Ti$ 体系 663K 真空退火 24h 固态反应周期层片型结构

**Fig.4** SEM images of periodic-layered structure in Zn/$Cu_3Ti_2$ diffusion couple annealed at 663k for 24 h

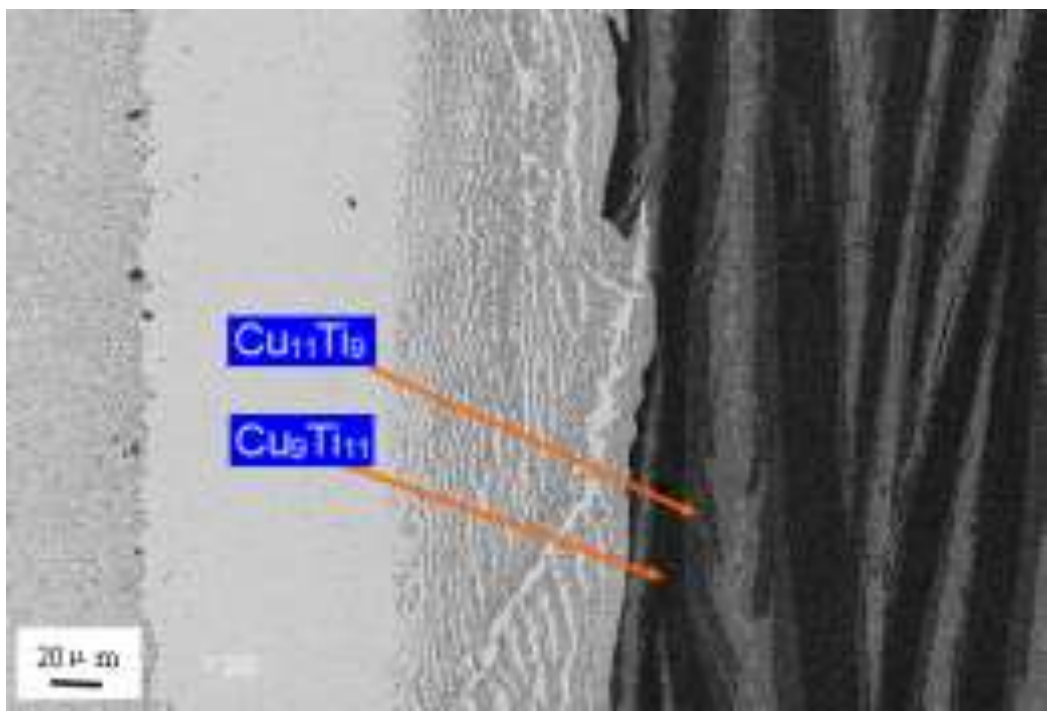

**图 5** Zn/$Cu_4Ti_3$ 体系 663K 真空退火 24h 固态反应周期层片型结构

**Fig.5** SEM images of periodic-layered structure in Zn/$Cu_4Ti_3$ diffusion couple annealed at 663k for 24 h

### 2.2 层状结构单双相的确定

由(α+β)双相构成的层片微观组织非常细腻，只有微米甚至纳米的分辨级别，因此在金相抛光后的扫描观察中，该双相层片往往容易被误认为是由单相β构成的，而不是双相的束集结构。有些体系，例如 $Mg/SiO_2$、$Zn/Ni_3Si$，在最初报道时就被误认为是单相α与单相β交替构成的周期层片结构[7,8]。

如图 3、图 4 所示，$Zn/Cu_xTi_y$ 体系周期型结构是由两种层片交替排列构成，但低倍率下不能确定周期排列的层片是单双相交替构成还是由两种单相构成。图 6 是对 $Zn/Cu_3Ti_2$ 体系周期层片型结构的观测照片，a 图放大倍数较低，给读者层状结构是由两种单相构成的假象，b 图放大倍数较大，可以看到层结构中明纹颜色大体一致，其上点缀的不连续的暗点为β相，其含量较少，则认为明纹为单相层；图中暗纹弥散着白色亮点，其均匀的分布状态不可能是金相制备引起，应该为层结构中包含的一种成分，由此可以推测暗纹并不是单相层。

对反应区的金相观测并没有很直观展现出层片的相结构，图 7 是对 Zn/CuTi 体系周期片层型结构断口原位观测的 SEM 图，图中能清楚地辨别出周期层片结构是单双相交替排列构成，其中单相层平整光滑连为一体，双相层中两相纠结在一起形成的束集型结构，此结果符合扩散应力模型的预测。

文献[24]认为周期性层片结构是两种单相交替出现，应该是由于其实验中只进行了金相观测且放大倍数不够，观测结果对作者产生了误导。

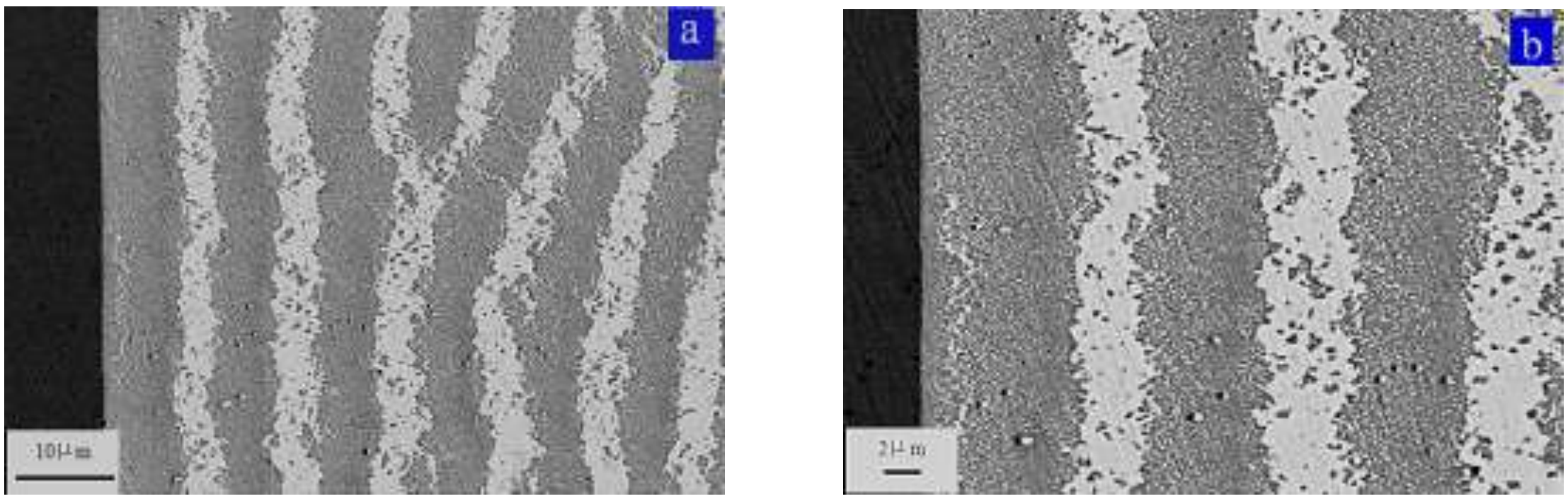

**图 6** $Zn/Cu_3Ti_2$ 体系 663K 真空退火 24h 后周期片层型结构高倍观测(金相观测）

**Fig.6** High magnification observation SEM images of periodic-layered structure in $Zn/Cu_3Ti_2$diffusion couple after annealing at 663k for 24 h(metallography)

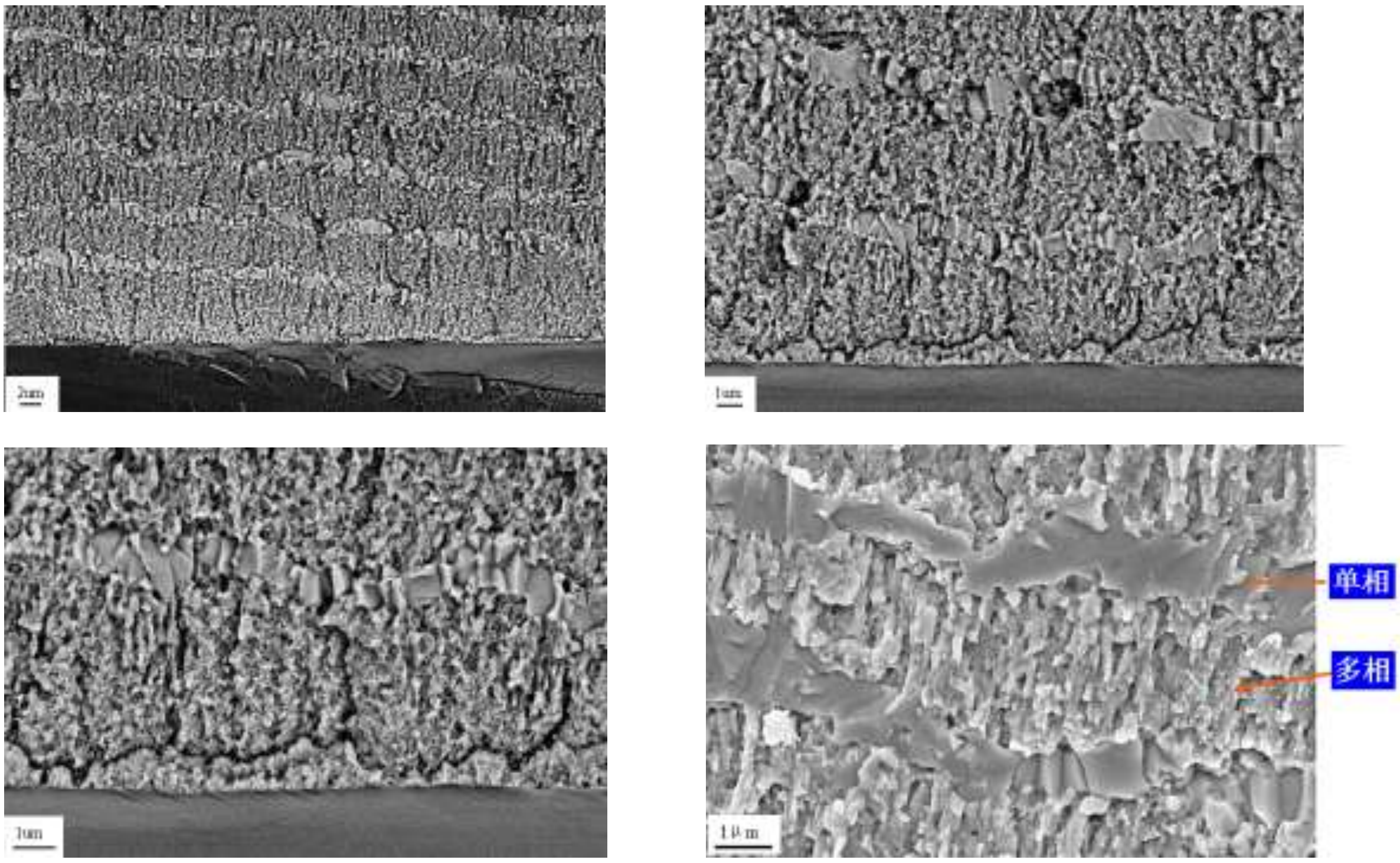

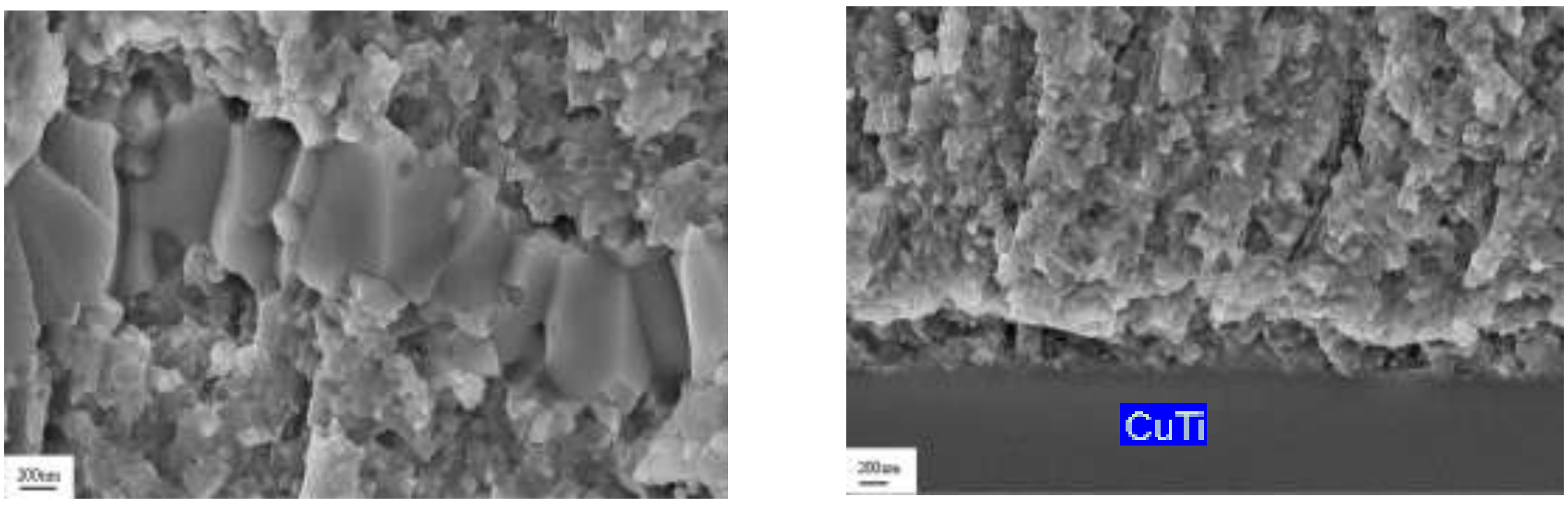

**图 7**　Zn/CuTi 体系 663K 真空退火 12h 后周期片层型结构高倍观测（断面观测）

**Fig.7** High magnification observation SEM images of periodic-layered structure in Zn/CuTi diffusion couple after annealing at 663k for 12h(section)

### 2. 3 层片厚度与合金成分的关系

如图 3、图 4、图 5 所示，成分不同的 Cu-Ti 合金与 Zn 金属反应生成的周期片层中双相层（暗纹）厚度不同，通过对不同反应体系层片厚度进行测量总结得到表 1，对比表中数据可以得出，随着合金中 Cu 原子含量的增加，双相层层片的厚度减小。

以 Zn/$Cu_3Ti_2$ 体系为例简述扩散应力模型。如图 8，左端为 Cu-Ti 基体来提供扩散反应的铜原子与钛原子，右端为锌基体提供锌原子，三种原子在扩散偶中的扩散速率大小依次为 Zn 原子最快，Cu 原子较慢，Ti 原子只进行局部的微扩散。当两种金属体接触，接触面发生扩散反应生成双相 $CuZn_2$（α 相）+$TiZn_3$（β 相）（$12Zn+Cu_3Ti_2{=}{=}3CuZn_2+2TiZn_3$），

**表 1** 不同反体系对应的双相层层片厚度

**Table 1** Thickness of layers in different solid reaction systems

| reaction systems | Thickness(μm) |
|---|---|
| Zn/CuTi | 5.38-7.50 |
| Zn/$Cu_3Ti_2$ | 3.56-4.88 |
| Zn/$Cu_7Ti_3$ | 1.40-1.70 |
| Zn/$Cu_4Ti$ | 1.30-1.48 |
| Zn/$Cu_9Ti$ | 0.85-1.10 |

双相层向着两边推进，Zn 原子扩散速率快，则 Cu-Ti 基体侧的反应前端 b 向左推进较快，Cu 原子和 Ti 原子扩散速率慢，则 Zn 基体侧反应前端推进较慢，

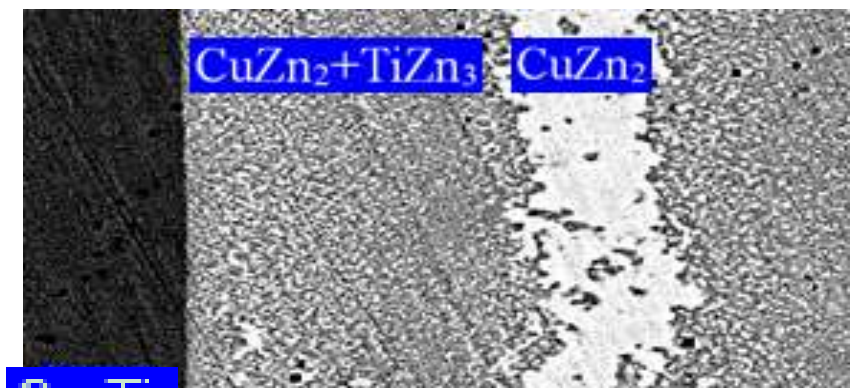

**图 8** 以 Zn/$Cu_3Ti_2$ 体系 663K 真空退火 48h 为例解释扩散应力模型

**Fig.8** Illustraton of the formation mechinism of periodic-layered structure in the Zn/$Cu_3Ti_2$ system annealed at 663k for 48h

当双相层生长到一定厚度，两种基体中各元素扩散到对方反应前端变的困难，由于 Ti 原子只能局部微移动，则 Cu 原子在 Zn 基体侧的反应前端积累的原来越多，形成了第一个单相 $CuZn_2$ 层，如 a；$CuZn_2$ 在 a 处生长的同时，反应前端 b 一直进行着扩散反应，当两相层厚度再次增大到一定程度后，Cu 原子扩散到 a 变得困难，此时两相区中的 $CuZn_2$ 重新开始生长，由于 $CuZn_2$:$TiZn_{3=}$3:2 的生成速率（$12Zn+Cu_3Ti_2{=}{=}3CuZn_2+2TiZn_3$）远大于稳定后双相中 $CuZn_2$:$TiZn_3$=9:19[14]的成分比例，使得生成速率快的 $CuZn_2$ 相对生长速率慢的 $TiZn_3$ 相产生一个内应力，当这个力积累到一定值，突破了 $TiZn_3$ 的强度极限使其发生撕裂，快速扩散的 Cu 和 Zn 原子填充到断裂处生成新的 $CuZn_2$ 层，如此重复，则形成了周期性层状结构。

由上述周期层片型结构形成过程的描述可知，随着 Cu-Ti 基体中 Cu 原子含量的升高，则基体中有更大的部分可以向反应区进行扩散，使得双向层中 α 相含量升高，更少的 β 相受到更大的积累应力，

则反应前端推进较窄的距离就使得 β 相撕裂，因此 Cu 原子含量越高，周期性层片越窄。

## 3 结论

（1）未经过均质化处理的 Cu-Ti 合金和 Zn 金属的扩散反应区中发现了 $Zn/Cu_9Ti$，$Zn/Cu_4Ti$，$Zn/Cu_2Ti$，$Zn/Cu_7Ti_3$，$Zn/Cu_3Ti_2$，$Zn/Cu_{11}Ti_9$，$Zn/Cu_9Ti_{11}$ 7 种未经报道的生成周期层片型结构的反应体系，这对新的反应体系的发现和复合材料的制备有启示意义。

（2）$Zn/Cu_xTi_y$ 体系中周期性结构中层片是按基体 Zn|双|单|双…单|双|Cu-Ti 基体的方式进行排列，金相抛光和断口原位观测都证实了双相的存在，符合扩散应力模型的预测。

（3）$Zn/Cu_xTi_y$ 体系中 Cu-Ti 基体成分会影响层片厚度，大致的规律是随着 Cu 含量的增加，层片厚度减小。

## 参考文献